\documentclass[a4paper,fleqn]{cas-sc}

\usepackage[authoryear,longnamesfirst]{natbib}
\usepackage{pgfplots}
\usepackage{booktabs}
\usepackage{float}
\usepackage{caption}

\pgfplotsset{compat=1.18}
\def\tsc#1{\csdef{#1}{\textsc{\lowercase{#1}}\xspace}}
\tsc{WGM}
\tsc{QE}

\begin{document}
\let\WriteBookmarks\relax
\def\floatpagepagefraction{1}
\def\textpagefraction{.001}

\shorttitle{Efficient domain-adaptation of Whisper on resource constrained devices}

\shortauthors{D. Koulouris, I. Maglogiannis}  

\title [mode = title]{Efficient domain-adaptation of Whisper on resource constrained devices}  

\tnotemark[1] 


%

\author[]{Dionysios Koulouris}[
    orcid=0000-0001-5325-6286,
]

\cormark[1]


\ead{dnkoulouris@unipi.gr}


\credit{Methodology, Software}

\affiliation[]{organization={University of Piraeus},
            addressline={Karaoli ke Dimitriou 80}, 
            city={Piraeus},
            postcode={18534}, 
            state={Attica},
            country={Greece}}

\author[]{Ilias Maglogiannis}[
    orcid=0000-0003-2860-399X,
]


\ead{imaglo@unipi.gr}


\credit{Conceptualization of this study, Supervision}



\cortext[1]{Corresponding author}



\begin{abstract}
Mobile speech recognition presents a persistent dilemma: rely on complex and expensive domain-specifically tuned cloud infrastructure or deploy generic on-device models that suffer from poor domain adaptation. The evolution of edge devices in terms of computational capabilities opens new avenues to bridge this gap. Achieving the high accuracy of resource-intensive cloud systems on constrained edge hardware drives the demand for algorithmic optimization, particularly through modern techniques that fine-tune foundational models for offline, on-device inference. In this work, we propose a novel pipeline for the domain-specific fine-tuning of foundational speech models for offline, on-device inference. Specifically, we apply Low-Rank Adaptation to the Whisper architecture, deploying and evaluating the optimized model on portable, resource-constrained hardware against real-world audio data. To prove the methodology robustness, datasets across various domains, languages and audio qualities were used. Results highlight that this on-device adaptation significantly improves speech recognition accuracy against baseline models. This research has a major impact mainly in high-stakes edge scenarios, offering critical advantages in strict data privacy, offline availability, reduced operational costs, and increased portability using commodity devices.
\nocite{*}
\end{abstract}


\maketitle

\section{Introduction}
Speech is the primary way of communication between humans, and, hence, methods by which speech is recognized are simultaneously a great need and a challenge. 
Computational speech recognition is an area that the literature is researching, with solutions including isolated and connected word along with continuous and spontaneous speech recognition \cite{gaikwad2010review}. 
From the early stages of studies, researchers faced challenges to identify the acoustic, phonetic, syntactic and semantic differentiations bound to a specific phrase \cite{reddy1976speech}. 
As technology evolved, and specifically Deep Neural Networks (DNNs), Automatic Speech Recognition (ASR) systems were introduced, allowing the conversion of spoken language into written text using computational algorithms \cite{yu2016automatic}.

Latest advancements include the introduction and combination of transformers technology and Convolutional Neural Network (CNN) into ASR solutions. 
CNN models are effectively exploiting features retrieved from transformer models, introducing new area for research in recognizing audio sequences in a parameter-efficient way \cite{gulati2020conformer}.
Transformers are generally enhancing accuracy and robustness of various models especially in challenging acoustic scenarios \cite{sharrab2025advancements}.

Although transformer-based ASR models achieve excellent performance on general vocabulary, due to their self-attention mechanisms, they frequently exhibit erroneous recognitions in scenarios requiring highly specialized terminology. 
In domain-specific scenarios, this strong internal language bias often causes overruling of the acoustic features of the model. 
Rare, out-of-vocabulary (OOV) technical terms are substituted with phonetically similar but contextually incorrect conversational words \cite{ye2022improving}. 
To counter these semantic hallucinations, domain adaptation techniques — like parameter-efficient fine-tuning (PEFT) — are utilized to move the focus of the model in the vector space towards the target vocabulary \cite{gupta2024parameter}.

However, a significant operational bottleneck appears when these adapted, large-scale foundational models are executed outside of server or cloud environments. 
Traditional cloud hardware is usually capable of absorbing the massive memory and computational overhead required for adapted transformers \cite{feng2025edge}. 
While efficient, this reliance is not accepted when stringent data privacy, high availability or network constraints demand local, offline execution. 
Consequently, deploying these compute-intensive, domain-adapted architectures on low-resource edge devices remains a critical challenge, establishing a distinct gap between cloud-level accuracy and on-device hardware limitations \cite{ramesh2023federated}.

In this work we propose a framework for efficient Whisper domain adaptation that can be transferred fully offline on commodity edge devices while retaining domain-specific accuracy.
Rather than presenting a standard domain adaptation methodology, the key contribution of this work is the targeted fine-tuning of ASR models, enabling offline inference on mobile and resource-constrained commodity devices. 
Beyond mitigating computational overhead and network latency, this edge-based architecture provides strict privacy preservation by performing all speech inference locally eliminating dangers related to cloud-based data transmission. 
Results show that the fine-tuned models outperform the baseline model in all tested scenarios, indicating that Low-Rank Adaptation (LoRA)-based fine tuning can be used as a generic pipeline for efficiently adapting ASR execution in resource-constrained devices to any domain. The rest of the paper is structured as follows: 

\section{Related Work}
This section reviews the evolution of acoustic modelling, transitioning from early deep learning approaches to modern parameter-efficient transformer architectures on the edge.

\subsection{General Speech Recognition and Deep Learning}
Acoustic modelling is a topic that concerns the scientific community and forms the foundation of speech recognition. 
In earlier stages, speech recognition was performed through Gaussian Mixture Models (GMM-HMM) with results demonstrating limited accuracy. 
The introduction of Deep Neural Networks (DNNs) formed a state-of-the-art shift as DNN multi hidden layer architecture outperforms GMMs on various benchmarks \cite{hinton2012deep}. 
More recent studies highlight that Connectionist Temporal Classification (CTC) methods, attention-based encoder-decoder models and Recurrent Neural Network Transducers (RNN-T) are scoring higher benchmarks and show better real-world deployment characteristics \cite{nayeem2025automatic}.

\subsection{Transformer-Based ASR and Foundational Models}
One of the latest revolutions in transduction models and in Artificial Intelligence in general is the introduction of attention through transformers \cite{vaswani2017attention}. 
Foundational self-attention mechanisms are underpinning all modern transformer-based ASR architectures. Studies show that self-supervised learning drastically reduces the need for annotated data, by accessing raw audio to learn latent speech representations. 
This method is implemented by wav2vec 2.0, a framework that demonstrates feasibility of speech recognition with limited training data \cite{baevski2020wav2vec}. 
In addition, the combination of transformers and CNNs enhances ASR accuracy by simultaneously capturing local acoustic features and global dependencies of the audio waveforms \cite{gulati2020conformer}. 

\subsection{State-of-the-art in ASR}
A self-supervised framework which achieves acoustic-representation learning is HuBERT. This system is predicting discrete cluster assignments and reduces the need for labeled ASR datasets drastically \cite{hsu2021hubert}.
By applying specific tuning, such models are capable of successfully recognizing large number of languages \cite{pratap2023scaling}, with wav2vec 2.0 extension and Google Universal Speech Model (USM) \cite{zhang2023google} being two paradigms of this self-supervised learning approach.
Furthermore, a unified, state-of-the-art foundational model, SeamlessM4T by Meta, performs well under multimodal and cross-lingual speech-to-text tasks \cite{barrault2023seamlessm4t} and pushes the boundaries of what an end-to-end architecture can achieve. 
Finally, by bypassing standard self-supervised learning in favor of 680,000 hours of weakly supervised internet audio yields unprecedented zero-shot generalization across varied acoustic environments \cite{radford2023robust}, introducing Whisper. 
This model achieves great performance across various accents and speaker traits \cite{graham2024evaluating}. 
An extensive benchmark explored among others the aforementioned models against Whisper and concludes that Whisper is the best and most robust architecture to handle real-world audio data, background noise and non-related-to-context voices \cite{shah2025speech}.

\subsection{PEFT and LoRA in ASR}
LoRA defines the fine-tuning of massive, large scale models by providing pre-trained weights and injecting trainable rank decomposition matrices \cite{hu2021lora}.
This technique is applied in various domains to direction the focus of a model towards a specific region in the vector space.
For instance, in Image Analysis, is used in synthetic image detection  \cite{keita2025bi}, in medical image parsing \cite{shaaban2025me} or in medical forensic image-to-text translation \cite{kallipolitis2026fine}.
The demand for robust domain adaptation in this field has driven various recent architectural innovations, ranging from compositional attention merging \cite{shankar2027compositional} and the evaluation of modern LLM re-scoring in highly specific linguistic contexts \cite{gonzalez2025modern}, to overcoming limited data scenarios through uncertainty-aware language model training \cite{sheikh2024training}.
Finally, speech is a domain that LoRA is successfully applied \cite{yu2023low}, with proposed prototypes achieving good results in various speech-to-text tasks.
Whisper \cite{song2024lora} is examined for its multilingual support and is performing well, even with limited training data \cite{xu2024towards}.

\subsection{Edge ASR Environments and The Scientific Gap}
While transformer-based ASR models have increased accuracy and efficiency, they require large amount of energy and computational resources. 
These operational trade-offs must be carefully resolved to enable local, transformer-based ASR inference, a capability that is strictly required to bypass the connectivity and latency bottlenecks of cloud-dependent architectures \cite{gondi2021performance}.
In scenarios where data privacy is operational requirement, related vulnerabilities inherent in sending sensitive, unlabeled audio from edge devices to centralized cloud servers for processing \cite{ramesh2023federated}.

The memory and computational needs of baseline ASR models lead to using model versions with fewer parameters which can be executed on portable devices. 
In their original form these models do not produce accepted results in their domain, hence the need of Low-Bit Quantization pipelines is present.
This is the scientific gap that this work is investigating, by fine-tuning a capable of running in commodity devices ASR model in various domains and languages, and comparing the results against baseline on-device inference using the same model.

\section{Methodology}
The primary objective of this research is to fulfil the gap between high-accuracy yet cloud-dependent speech recognition and the constraints of resources regarding edge-based inference. 
To eliminate the prohibitive computational cost of retraining massive foundational models, a targeted domain adaptation strategy is formulated.
Specifically, LoRA is utilized to inject trainable rank decomposition matrices into the transformer layers of a pre-trained baseline ASR architecture. 
This pipeline shifts the focus of the model in the vector space towards targeted domain-specific vocabulary, maintaining the integrity and general linguistic knowledge of the foundational model.
By adopting this modular approach, the system allows for the dynamic contextualization of the model during inference, striking an optimal balance between the memory limitations of commodity hardware and the necessity for state-of-the-art accuracy in privacy-sensitive, offline scenarios.

This section details the approach taken to develop and deploy highly specialized, privacy-preserving ASR systems for edge devices. 
The methodology is structured into two core components: the system architecture and the technical implementation.

\subsection{Architecture}
\begin{figure}
  \centering
   \includegraphics[width=0.8\textwidth]{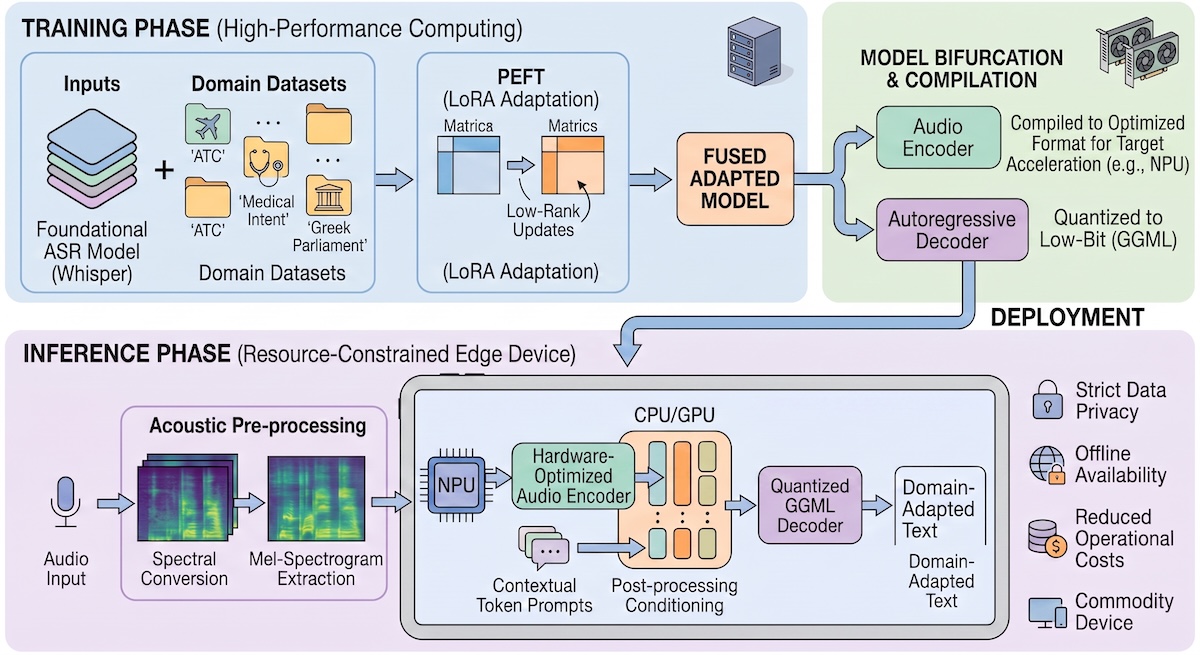}
    \caption{
    The pipeline consists of three phases, the training phase, the compilation phase and the inference phase.
    The training phase retrieves the foundational model and the datasets, performs processing and LoRA before outputting a domain-adapted model.
    This model is inserted to the compilation phase, where it is compiled to device-optimized executable.
    The inference phase is executed offline, on the edge device. 
    The audio input in addition with contextual token prompts is passed through acoustic pre-processing and encoding, post-processing conditioning and quantized decoding. 
    The final domain-adapted text output is extracted.
    }\label{fig1}
\end{figure}

The proposed system architecture is underpinned by a robust transformer-based ASR foundational model, specifically selected for its zero-shot generalization capabilities and inherent parameter efficiency. 
To enable local, offline inference on edge devices, the architecture leverages a hybrid execution strategy that rigorously exploits on-device hardware accelerators.
It consists of three phases: the training phase, the compilation phase and the inference phase.
The computationally dense audio encoder component of the transformer is compiled into an optimized format designed to be aggressively offloaded to dedicated Neural Processing Units (NPUs). 
This offloading ensures high-throughput, low-power processing of acoustic features, primarily the extraction of mel-spectrograms. 
Conversely, the autoregressive decoder operates within a highly optimized tensor library framework, utilizing low-bit quantization techniques to minimize memory bandwidth requirements while executing on the Central or Graphics Processing Units. 
This hybrid architectural approach effectively mitigates the latency and computational bottlenecks traditionally associated with transformer-based ASR inference on portable hardware. The architecture of the system is shown in Figure \ref{fig1}.

\subsection{Implementation}
The practical implementation of the proposed methodology encompasses a strict pipeline designed to bridge high-performance domain adaptation and resource-constrained edge execution. 
This process is functionally categorized into three core operational phases.

\subsubsection{Training phase}
This initial phase focuses on tailoring the foundational ASR architecture to specialized vocabularies without retraining the entire massive network from scratch. 
Acoustic datasets are rigorously pre-processed to align with the strict input tensor dimensions required by the foundational model.
During adaptation, domain-specific LoRA weights are mathematically optimized against these pre-processed acoustic datasets within a high-performance computing environment.
Specifically, the fine-tuning phase selectively targeted the self-attention projection modules of the foundational transformer, utilizing a LoRA rank ($r$) of 32 and a scaling factor ($\alpha$) of 64 to ensure robust gradient updates without destabilizing the pre-trained semantic space. 
Optimization was driven by the AdamW algorithm coupled with a linear learning rate decay schedule, evaluating cross-entropy loss across the autoregressive token generation.
Post-training, the newly optimized low-rank adapter matrices are statically fused with the pre-trained baseline foundational weights. This yields a unified, highly specialized model topology ready for optimization.

\subsubsection{Compilation phase}
To achieve efficient edge deployment, the fused model must be aggressively optimized to respect the strict memory and thermal constraints of portable commodity hardware.
The unified model is structurally separated into its two primary components: the audio encoder and the autoregressive decoder.
The computationally dense audio encoder is heavily optimized and compiled into a format designed specifically for hardware-level neural acceleration.
This ensures high-throughput, low-power processing on dedicated Neural Processing Units.
The autoregressive decoder undergoes low-bit quantization, a process that drastically reduces its required memory footprint and bandwidth, enabling it to execute efficiently within a standard CPU or GPU framework.

\subsubsection{Inference phase}
The final phase dictates how the deployed, optimized model handles real-time, offline speech recognition entirely on the edge device.
Raw audio waveforms captured by the edge device microphone undergo immediate structural standardization.
They are aggressively resampled to a uniform 16kHz rate (satisfying the Nyquist-Shannon sampling theorem), down-mixed to a monaural channel, and normalized to a 16-bit Pulse-Code Modulation (PCM) representation.
Overlapping windowed segments then undergo a discrete Fourier transform to extract highly compressed, perception-aligned Mel-spectrograms for the neural encoder.
Before decoding begins, the process is heavily regulated by injecting deterministically structured contextual tokens.
This artificially biases the prior probability distribution of the language model toward the expected domain vernacular.
As the model translates latent neural representations into text, the initial prompt conditioning acts as a strict post-processing filter.
It effectively suppresses generic conversational vocabulary in favor of out-of-vocabulary (OOV) technical terms and local dialects, systematically mitigating semantic hallucinations.

\section{Results}
\subsection{Evaluation methodology}
To validate the proposed domain-adaptation pipeline and on-device inference architecture, a comprehensive experimental framework was established. 
This subsection details the hardware environments utilized for both training and edge deployment, the diverse datasets selected to evaluate domain and cross-lingual robustness, and the specific metrics employed to rigorously quantify the results.

\subsubsection{Hardware}
The parameter-efficient fine-tuning phase of the foundational model was executed within a high-performance computing environment utilizing CUDA-accelerated Graphical Processing Units (GPUs). 
Conversely, the proposed inference pipeline is fundamentally engineered to execute efficiently across any high-performance mobile commodity devices. 
The validation of the paradigm was conducted on an Apple Silicon System-on-Chip (SoC) equipped with a dedicated Neural Processing Unit, enabling strict offline execution of the compiled models.
It should be explicitly noted that this specific hardware configuration was utilized solely for reference purposes to benchmark the methodology, as the proposed domain adaptation pipeline remains inherently platform-agnostic.

\subsubsection{Dataset Selection}
To prove the robustness of the methodology against varying acoustic environments, languages, and semantic domains, four distinct evaluation datasets were utilized. Each dataset presents unique computational challenges for the foundational ASR model:
\begin{enumerate}[a)]
\item Air Traffic Control (ATC)\footnote{\url{https://huggingface.co/datasets/jacktol/ATC-ASR-Dataset}}: Characterized by high background noise, extreme radio compression, and heavy reliance on aviation-specific jargon. This dataset severely tests the model's ability to suppress generic conversational vocabulary in favor of out-of-vocabulary (OOV) technical phrasing.
\item Medical Intent\footnote{\url{https://huggingface.co/datasets/ttthe/MultiMed}, English Configuration}: A highly specialized dataset where strict morphological precision is critical. Misinterpretation of complex pharmacological or anatomical terms can lead to catastrophic semantic failures, demanding high character-level accuracy.
\item Greek Parliament\footnote{\url{ https://huggingface.co/datasets/Elormiden/Hellenic-greek-parliamentary-speech}}: A cross-lingual dataset featuring formal, structured political discourse. It evaluates the model's ability to retain the foundational architecture's inherent multilingual capabilities while adapting to specific national vernaculars.
\item German Speech\footnote{\url{https://huggingface.co/datasets/ttthe/MultiMed}, German Configuration}: A cross-lingual general speech dataset utilized as a control benchmark. It ensures that the domain-specific adaptation pipeline does not degrade the model's baseline performance on standard conversational audio in non-English languages.
\end{enumerate}  

\subsubsection{Evaluation metrics}
A multidimensional evaluation criteria was established to rigorously quantify the efficacy of the on-device inference. Relying solely on a singular metric provides an incomplete assessment of ASR performance, particularly in highly specialized domains.
\begin{itemize}
\item Word Error Rate (WER): Utilized as the standard baseline metric to measure the raw frequency of word-level insertions, deletions, and substitutions.
\item Character Error Rate (CER): Employed to assess strict morphological precision. CER is highly sensitive to spelling anomalies, making it critical for evaluating performance on the Medical Intent dataset, where a single incorrect character drastically alters semantic meaning.
\item BLEU Score: A standard machine translation metric applied to evaluate n-gram structural fluency. It measures the precision of the hypothesized text against the reference transcripts, ensuring that domain-specific phrasing (e.g., ATC jargon) is sequentially accurate.
\item Semantic (Cosine) Similarity: To definitively quantify deep semantic equivalence, both the generated hypotheses and the ground truth were mapped into a 384-dimensional vector space using a state-of-the-art multilingual transformer architecture (paraphrase-multilingual-MiniLM-L12-v2). The cosine similarity of these dense embeddings was calculated to ensure that the exact semantic intent of the speaker was preserved, completely independent of minor lexical or grammatical deviations.
\end{itemize}  

\subsection{Quantitative results}
This section presents a comprehensive quantitative analysis evaluating the efficacy of the proposed domain-adaptation pipeline. 

To rigorously assess the performance gains, the fine-tuned on-device models are benchmarked directly against their foundational baseline counterparts across all four acoustic datasets. 
The evaluation is conducted utilizing a multidimensional metric framework—encompassing Semantic Cosine Similarity, Character Error Rate (CER), and BLEU scores—to capture improvements in semantic fidelity, morphological precision, and n-gram structural fluency, respectively.
The comparative outcomes of these evaluations are detailed in the subsequent tables and visually delineated through the accompanying performance graphs, illustrating the significant impact of the targeted fine-tuning methodology on edge-based inference.

\subsubsection{Testing corpus}
Table \ref{tab:corpus_details} outlines the domains, languages, and exact sample sizes used for evaluation based on CSV data.
\begin{table}[ht]
\centering
\caption{Overview of the diverse acoustic corpora utilized for the quantitative evaluation. The datasets were specifically selected to rigorously benchmark the domain adaptation pipeline across highly specialized, zero-tolerance environments (Medical Intent, Air Traffic Control) and cross-lingual structured political discourse (Greek Parliament), utilizing a standard conversational dataset (German Speech) as a non-English control benchmark. The table details the semantic domain, primary language, and total sample size of each corpus.}
\label{tab:corpus_details}
\begin{tabular}{l l l c}
\toprule
\textbf{Dataset} & \textbf{Domain} & \textbf{Language} & \textbf{Size (Samples)} \\
\midrule
Medical Intent   & Medical / Pharmacology & English & 500 \\
Air Traffic Control & Aviation / Radio Comm. & English & 1000 \\
Greek Parliament & Political / Formal Speech & Greek & 1000 \\
German Speech    & General Conversational & German & 200 \\
\bottomrule
\end{tabular}
\end{table}

\subsubsection{Word Error Rate}
WER is the foundational baseline metric for evaluating ASR algorithms.
It is calculated to assess the raw frequency of word-level inaccuracies, as shown in Table \ref{tab:wer}.
The application of the proposed pipeline resulted in significant WER reductions across three of the four specialized domains, most notably yielding a 61.79\% absolute reduction in the ATC dataset.
A statistically negligible regression (+0.34\%) was observed in the German Speech control dataset.
However, this behaviour is mathematically consistent with domain adaptation, due to biasing the probability distribution toward specialized technical vernaculars. 
Overall, WER improvements validate that the proposed architecture suppresses the broad hallucinations in favour of strict domain compliance.

\begin{table}[h]
\centering
\caption{Word Error Rate (WER) percentages across the evaluated datasets. 
Lower scores represent higher word-level accuracy. 
The data highlights a substantial reduction in word-level insertions, deletions and substitutions for the proposed model, particularly in the highly specialized Air Traffic Control and Medical Intent domains. 
The marginal regression in the German Speech control corpus confirms that aggressive domain adaptation inherently biases the decoder away from general conversational structures.}
\label{tab:wer}
\begin{tabular}{l c c c}
\toprule
\textbf{Dataset} & \textbf{Original WER} & \textbf{Proposed WER} & \textbf{Absolute Difference} \\
\midrule
German Speech    & \textbf{11.16\%} & 11.50\% & +0.34\% \\
Medical Intent   & 31.35\% & \textbf{21.08\%} & -10.27\% \\
ATC Dataset      & 77.66\% & \textbf{15.87\%} & -61.79\% \\
Greek Parliament & 22.04\% & \textbf{16.79\%} & -5.25\% \\
\bottomrule
\end{tabular}
\end{table}

\subsubsection{Character Error Rate}

\begin{table}[h]
\centering
\caption{Character Error Rate (CER) percentages across the evaluated datasets. Lower scores represent higher morphological precision. Unlike standard word-level metrics, CER is highly sensitive to spelling anomalies, making it a critical benchmark for highly specialized environments (such as the Medical Intent corpus) where a single incorrect character can catastrophically alter the semantic meaning of complex terminology. The data illustrates that the proposed domain adaptation yields a drastic reduction in character-level errors, proving that the fine-tuned autoregressive decoder accurately predicts complex, domain-specific lexical structures with near-perfect morphological fidelity compared to the baseline.}
\label{tab:cer}
\begin{tabular}{l c c c}
\toprule
\textbf{Dataset} & \textbf{Original CER} & \textbf{Proposed CER} & \textbf{Absolute Reduction} \\
\midrule
German Speech    & 9.69\% & \textbf{9.22\%} & -0.47\% \\
Medical Intent   & 30.77\% & \textbf{23.49\%} & -7.28\% \\
ATC Dataset      & 83.02\% & \textbf{8.62\%} & -74.40\% \\
Greek Parliament & 23.39\% & \textbf{14.81\%} & -8.59\% \\
\bottomrule
\end{tabular}
\end{table}

As detailed in Table \ref{tab:cer}, the proposed architecture yields a drastic reduction in Character Error Rate compared to the baseline, proving that the targeted fine-tuning successfully calibrates the decoder to accurately predict complex, domain-specific lexical structures down to the exact morphological level.
The comparison is demonstrated in Figure \ref{fig:cer_chart}.

\begin{center}
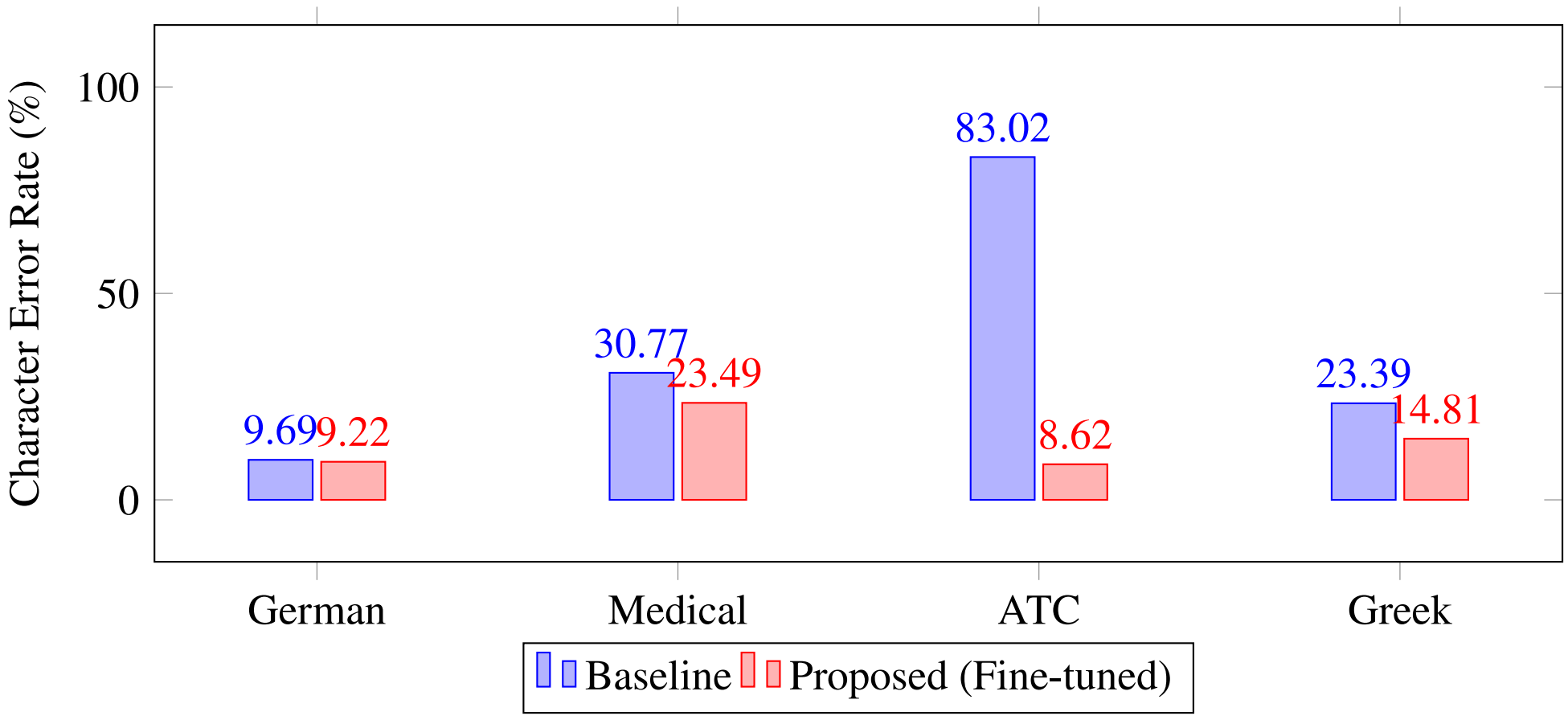

\begin{tikzpicture}
\begin{axis}[
    ybar,
    bar width=15pt,
    enlargelimits=0.15,
    ylabel={Character Error Rate (\%)},
    symbolic x coords={German, Medical, ATC, Greek},
    xtick=data,
    nodes near coords,
    nodes near coords align={vertical},
    legend style={at={(0.5,-0.15)}, anchor=north, legend columns=-1},
    ymin=0, ymax=100,
    width=0.8\textwidth,
    height=6cm
]
\addplot coordinates {(German,9.69) (Medical,30.77) (ATC,83.02) (Greek,23.39)};
\addplot coordinates {(German,9.22) (Medical,23.49) (ATC,8.62) (Greek,14.81)};
\legend{Baseline, Proposed (Fine-tuned)}
\end{axis}
\end{tikzpicture}
\captionof{figure}{Character Error Rate (CER) comparison. Lower values indicate better morphological precision.}
\label{fig:cer_chart}
\end{center}

\subsubsection{BLEU score}

\begin{table*}[h]
\centering
\caption{BLEU scores across the evaluated datasets. Higher scores indicate superior n-gram structural fluency and domain phrasing. The BLEU metric rigorously evaluates the precise sequential overlap between the generated transcription and the reference ground truth. The results confirm that the parameter-efficient domain adaptation pipeline successfully enables the foundational architecture to recognize and correctly sequence highly specialized phrasing---most notably in rigidly templated environments such as Air Traffic Control (+75.11 absolute improvement)---which the zero-shot baseline model inherently misinterprets as unstructured conversational noise.}
\label{tab:bleu}
\begin{tabular}{l c c c}
\toprule
\textbf{Dataset} & \textbf{Original BLEU} & \textbf{Proposed BLEU} & \textbf{Absolute Improvement} \\
\midrule
German Speech    & 65.90 & \textbf{69.21} & +3.31 \\
Medical Intent   & 53.33 & \textbf{67.26} & +13.93 \\
ATC Dataset      & 2.11 & \textbf{77.22} & +75.11 \\
Greek Parliament & 43.43 & \textbf{72.79} & +29.36 \\
\bottomrule
\end{tabular}
\end{table*}

Table \ref{tab:bleu} illustrates the superior n-gram structural fluency of the proposed model over the baseline, confirming that the domain adaptation pipeline enables the architecture to correctly sequence highly specialized phrasing that the baseline model inherently misinterprets as conversational noise.
The comparison is shown in Figure \ref{fig:bleu_chart}.

\begin{center}
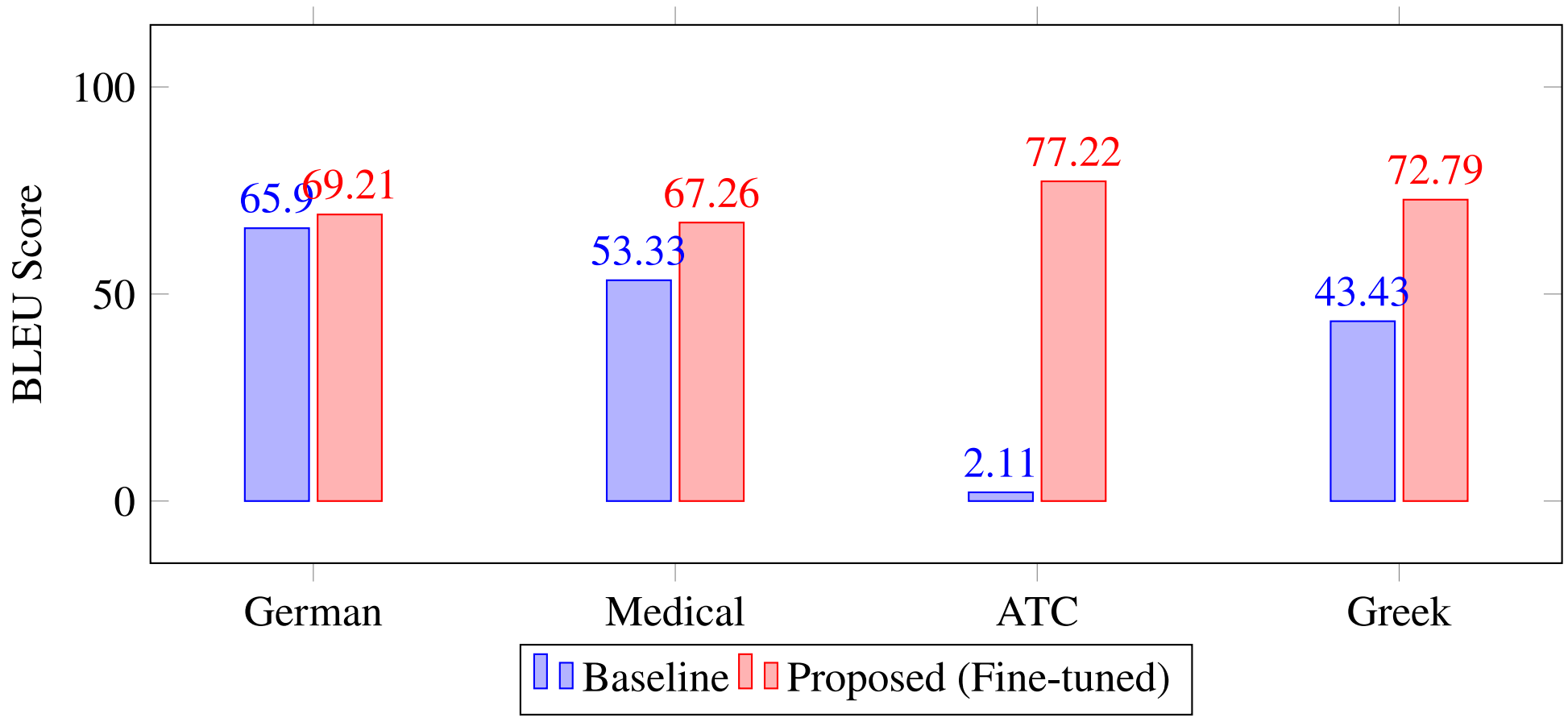

\begin{tikzpicture}
\begin{axis}[
    ybar,
    bar width=15pt,
    enlargelimits=0.15,
    ylabel={BLEU Score},
    symbolic x coords={German, Medical, ATC, Greek},
    xtick=data,
    nodes near coords,
    nodes near coords align={vertical},
    legend style={at={(0.5,-0.15)}, anchor=north, legend columns=-1},
    ymin=0, ymax=100,
    width=0.8\textwidth,
    height=6cm
]
\addplot coordinates {(German,65.90) (Medical,53.33) (ATC,2.11) (Greek,43.43)};
\addplot coordinates {(German,69.21) (Medical,67.26) (ATC,77.22) (Greek,72.79)};
\legend{Baseline, Proposed (Fine-tuned)}
\end{axis}
\end{tikzpicture}
\captionof{figure}{BLEU score comparison. Higher values indicate superior n-gram structural fluency.}
\label{fig:bleu_chart}
\end{center}

\subsubsection{Semantic (Cosine) Similarity}
\begin{table}[h]
\centering
\caption{Semantic (Cosine) Similarity across the evaluated datasets. To quantify deep semantic equivalence, both the generated hypotheses and the ground truth were mapped into a high-dimensional vector space utilizing a state-of-the-art multilingual transformer architecture. The cosine similarity of these dense embeddings, ranging from -1 to 1, mathematically proves that the proposed fine-tuning pipeline effectively suppresses out-of-vocabulary hallucinations and accurately preserves the core semantic intent of the speaker, significantly outperforming the zero-shot baseline across all tested domains.}
\label{tab:semantic_similarity}
\begin{tabular}{l c c c}
\toprule
\textbf{Dataset} & \textbf{Original (Baseline)} & \textbf{Proposed (Fine-tuned)} & \textbf{Absolute Improvement} \\
\midrule
German Speech    & 0.9344 & \textbf{0.9476} & +0.0133 \\
Medical Intent   & 0.7602 & \textbf{0.8737} & +0.1135 \\
ATC Dataset      & 0.5103 & \textbf{0.9150} & +0.4048 \\
Greek Parliament & 0.7125 & \textbf{0.8584} & +0.1458 \\
\bottomrule
\end{tabular}
\end{table}

Table \ref{tab:semantic_similarity} demonstrates that the proposed solution significantly outperforms the baseline by effectively suppressing out-of-vocabulary hallucinations, thereby shifting the model's language bias to mathematically preserve the core semantic intent of the highly specialized domains. The comparison is shown in Figure \ref{fig:semantic_chart}.

\begin{center}
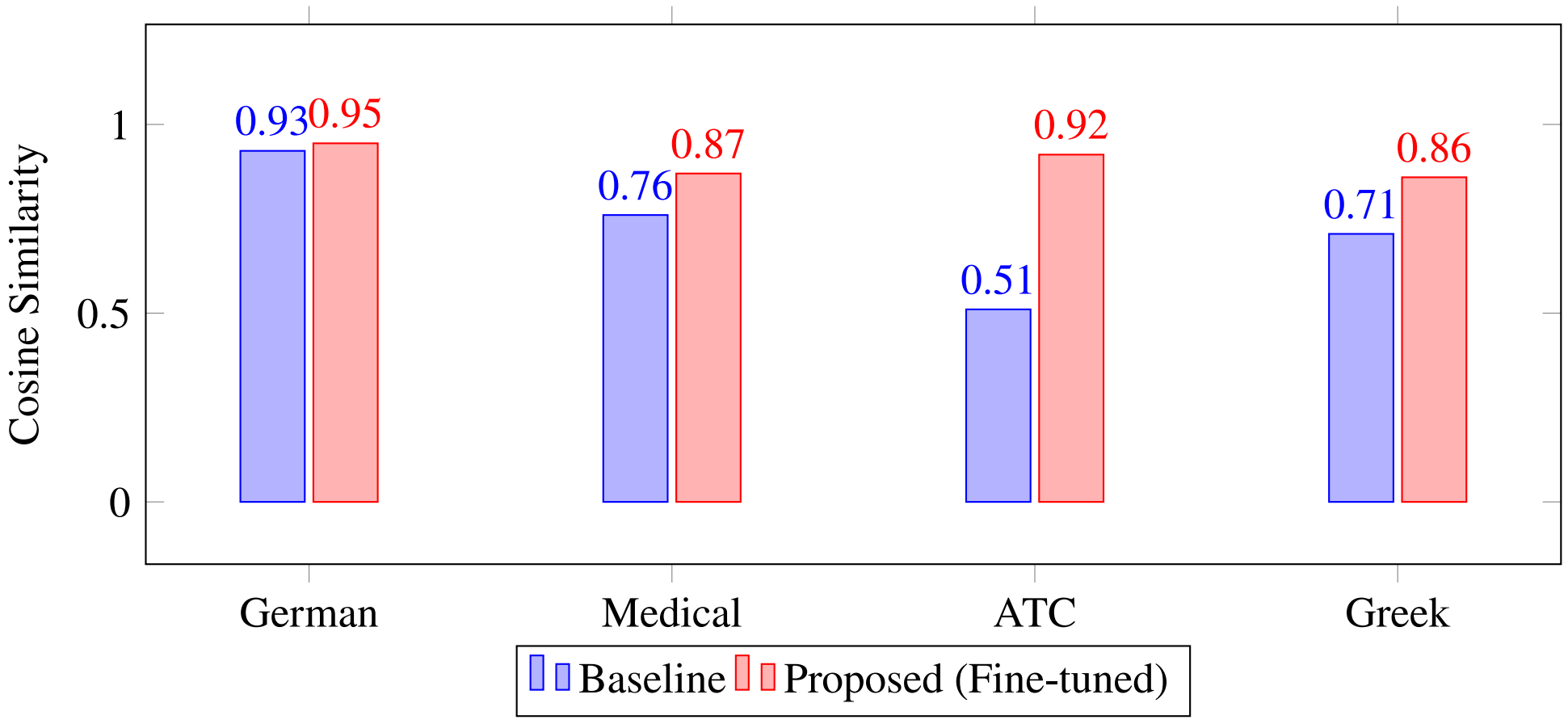

\begin{tikzpicture}
\begin{axis}[
    ybar,
    bar width=15pt,
    enlargelimits=0.15,
    ylabel={Cosine Similarity},
    symbolic x coords={German, Medical, ATC, Greek},
    xtick=data,
    nodes near coords,
    nodes near coords align={vertical},
    legend style={at={(0.5,-0.15)}, anchor=north, legend columns=-1},
    ymin=0, ymax=1.1,
    width=0.8\textwidth,
    height=6cm
]
\addplot coordinates {(German,0.93) (Medical,0.76) (ATC,0.51) (Greek,0.71)};
\addplot coordinates {(German,0.95) (Medical,0.87) (ATC,0.92) (Greek,0.86)};
\legend{Baseline, Proposed (Fine-tuned)}
\end{axis}
\end{tikzpicture}
\captionof{figure}{Semantic (Cosine) Similarity comparison between the Baseline and Proposed models. Higher is better.}
\label{fig:semantic_chart}
\end{center}

\subsubsection{Performance metrics}

Beyond evaluating transcription accuracy, the latency of the on-device execution was rigorously profiled to ensure the domain adaptation did not introduce operational bottlenecks. 
As detailed in Table \ref{tab:inference_time}, the average inference time per audio sequence remained effectively negligible, consistently executing in under one second across the majority of datasets. Notably, the proposed model executed faster than the baseline in the German Speech and Greek Parliament corpora. 

\begin{table}[h]
\centering
\caption{Average on-device inference latency per audio sample (in seconds). Lower values represent faster execution.}
\label{tab:inference_time}
\begin{tabular}{l c c c}
\toprule
\textbf{Dataset} & \textbf{Baseline Time (s)} & \textbf{Proposed Time (s)} & \textbf{Latency Difference (s)} \\
\midrule
German Speech    & 1.353 & \textbf{0.962} & -0.391 \\
Medical Intent   & \textbf{0.664} & 0.813 & +0.149 \\
ATC Dataset      & \textbf{0.775} & 0.896 & +0.121 \\
Greek Parliament & 0.799 & \textbf{0.747} & -0.052 \\
\bottomrule
\end{tabular}
\end{table}

\section{Discussion}
Domain adaptation in ASR is a highly studied topic with most contributions focusing on the optimization of recognition algorithms using high-performance computers.
Two major deficiencies are present when trying to achieve domain-specific ASR in mobile environments: the lack of computational power of mobile commodity devices and the continuous requirement for interconnection with a computer capable of executing such methods.
The goal of the proposed pipeline is to utilize a baseline ASR model---with a size capable of being executed on mobile commodity devices---and perform the correct adaptation to achieve better speech recognition.
By implementing a LoRA fine-tuning in Whisper, in various languages and domains, this work highlights that it can successfully bridge the gap between cloud-level accuracy and edge-device constraints.
Across all tested domains, the proposed model consistently outperformed the zero-shot foundational baseline without requiring a massive increase in on-device memory footprint.

A primary aspect of this approach is the preservation of semantic fidelity in high-stakes environments.
For instance, mitigating morphological errors (such as misspelling a drug name or an airline number) is critical in medical/pharmacological or transportation scenarios.
As retrieved from the CER results, our fine-tuning successfully suppressed these dangerous out-of-vocabulary hallucinations, offering more reliable speech recognition.

This study shows that the proposed approach has its highest impact in templated speech. 
As shown in all results of the ATC dataset, and specifically in the high BLEU score, our pipeline is highly capable of tuning models to perform ASR on strict, legally mandated phraseology.
While the baseline fails when heavy radio noise and jargon are present, the fine-tuned model easily locked onto the n-gram templates, proving highly effective for rigidly structured environments.

The proposed pipeline maintains its cross-lingual accuracy, as it is tested across three languages, with different structure, syntax, grammar, punctuation and peculiarities in pronunciation.
The foundational model's inherent multilingual capabilities were preserved, and the targeted adaptation successfully mapped to foreign national vernaculars and formal political discourse.

Apart from the technical and quantitative improvements of this pipeline, a benefit of the proposed solution is its independence from data exchange with other systems.
This fundamentally reduces operational overhead by eliminating the need to provision and maintain dedicated, always-on inference servers, and the need of a mobile internet connection, enhancing privacy.
By fine-tuning an ASR model and executing it in offline, mobile environments, it significantly mitigates the risk of data exfiltration and guarantees localized data retention.
Furthermore, executing models entirely on the edge eliminates reliance on local network infrastructure, which remains vulnerable to interference, latency, or downtime.
Our solution enables zero-trust, highly secure transcriptions for various scenarios (such as hospitals or governmental confidential discussions), without the latency or security vulnerabilities of transmitting audio to a centralized cloud.

As the model weights are decoupled from the base model, the proposed approach provides a high degree of domain flexibility, allowing different weight files to be loaded according to the target application scenario. 
Furthermore, various LoRA weight sets can be installed simultaneously on the mobile device, enabling the user to select the appropriate model at inference time. 
This design avoids locking the device to a single domain while supporting dynamic switching between multiple domain-specific models.

The latency reduction, although not extremely high, is mathematically consistent with autoregressive decoding behaviour of ASR models; by artificially biasing the probability distribution toward the correct vernacular, the fine-tuned model eliminates the computational overhead of generating prolonged, out-of-vocabulary semantic hallucinations. 
Ultimately, the data confirms that the integration of LoRA adapter matrices achieves state-of-the-art accuracy gains with zero penalization to edge-inference latency.

Regarding the limitations of the current study, battery consumption and thermal throttling are the two primary deficiencies. 
During empirical testing, battery capacity depleted at an approximate rate of 10\% per 500 inference tasks, but future research will focus on quantifying the exact joule-per-transcription energy overhead of the hybrid ANE/GGML pipeline on battery-constrained devices.
Additionally, thermal throttling presents a potential operational constraint; however, its onset is highly variable and strictly dependent on ambient environmental conditions.
Furthermore, a limitation is the time required for the initialization of the pipeline, as it might take up to a minute to fully load the adapted model in the system memory.

\section{Conclusions}
The proposed work presents a LoRA pipeline for ASR on resource-constrained commodity devices.
Key contribution of this work is the fine-tuning of low-parameter baseline ASR models for domain-specific inference in mobile devices. 
Various datasets were tested, including different languages, domains and ambient audio conditions.
Results show that, especially in templated vocabulary scenarios, our pipeline outperforms the baseline model both in accuracy, error reduction and inference time.
For future work, the dynamic real-time swapping of LoRA adapters based on acoustic environment detection can be examined, in combination with further quantizing the training (INT4 or INT8) for even smaller devices.
Moreover, additional domains and audio conditions can be tested, as well as real time tests can be performed.
Overall, this work demonstrates that the formidable capabilities of massive, foundational speech models can be successfully localized and fine-tuned for resource-constrained environments without incurring prohibitive latency. 
By coupling parameter-efficient domain adaptation with hybrid hardware acceleration, this methodology eliminates the historical compromise between strict data privacy and state-of-the-art semantic accuracy, establishing a highly robust framework for the next generation of autonomous, zero-trust edge intelligence.










\printcredits

\bibliographystyle{cas-model2-names}

\bibliography{cas-refs}



\end{document}